\documentclass[11pt,twoside]{article}
\usepackage{asp2010}

\resetcounters

\begin{document}

\title{The Distribution of Disk-Halo HI Clouds in the Inner Milky Way}
\author{H.~Alyson~Ford$^{1,2,3}$, Felix~J.~Lockman$^4$, and N.~M.~McClure-Griffiths$^2$
\affil{$^1$Centre for Astrophysics and Supercomputing, Swinburne University of Technology, Hawthorn, Victoria 3122, Australia;}
\affil{$^2$Australia Telescope National Facility, CSIRO Astronomy \& Space Science, Epping, NSW 1710, Australia;}
\affil{$^3$Department of Astronomy, University of Michigan, Ann Arbor, MI 48109, USA; haford@umich.edu}
\affil{$^4$National Radio Astronomy Observatory, Green Bank, WV 24944, USA.}}

\begin{abstract}
Using data from the Galactic All-Sky Survey, we have compared the properties and 
distribution of HI clouds in the disk-halo transition at the tangent points in mirror-symmetric
regions of the first quadrant (QI) and fourth quadrant (QIV) of the Milky Way. 
Individual clouds are found to have identical properties in the two quadrants.
However, there are 3 times as many clouds in QI as in QIV, their scale height is twice as large, and their
radial distribution is more uniform. We attribute 
these major asymmetries to the formation of the clouds in the spiral arms of the Galaxy, and suggest 
that the clouds are related to star formation in the form of gas that has been lifted from the disk by 
superbubbles and stellar feedback, and fragments of shells that are falling back to the plane.
\end{abstract}

\section{Introduction}

Early observations of the disk-halo interface hinted at the presence of a handful of HI clouds 
connected to the Galactic disk \citep{1964Prata, 1971Simonson, 1984Lockman}. With higher resolution data
it became evident that there is a population of discrete HI clouds within this transition zone of
the inner Galaxy, having sizes $\sim 30$~pc, masses $\sim 50$~$M_\odot$, and whose kinematics are dominated 
by Galactic rotation \citep{2002Lockman}. The presence of these clouds is a widespread phenomenon; 
they have also been detected in the disk \citep{2006Stil} and outer Galaxy \citep{2006Stanimirovic, 2010Dedes}.
These clouds may constitute the HI layer, and may be
detected in external galaxies once resolution requirements can be met. Possible origin scenarios
include galactic fountains (\citealt{1976Shapiro,1980Bregman,1990Houck}; \citealt*{2008Spitoni}),
superbubbles \citep{2006McClure-Griffiths}, and interstellar turbulence \citep{2005Audit}.

An analysis of the HI disk-halo cloud population in the fourth Galactic quadrant of longitude 
(QIV) was performed by \citet{2008Ford}. To study the variation of cloud properties and 
distributions with location in the Galaxy, \citeauthor*{2010Ford} (\citeyear{2010Ford}; hereafter
FLMG) analyzed a region in 
the first quadrant (QI) that is mirror-symmetric about longitude zero to the QIV region, 
and performed an in-depth comparison of the uniformly selected QI and QIV samples. In this 
proceeding we summarize some of the main results from that QI--QIV analysis. For an extended 
and complete discussion of these results we refer the reader to FLMG.

\section{Observations}

Disk-halo clouds were detected using data from the Galactic All-Sky Survey 
\citep[GASS;][]{2009McClure-Griffiths}, which were taken with the 21~cm Multibeam receiver at the Parkes
Radio Telescope. GASS is a Nyquist-sampled survey of Milky Way HI emission ($-400 \leq 
V_{\mathrm{LSR}} \leq +500$~km~s$^{-1}$), covering the entire sky south of $\delta \leq 1^\circ$.
The spatial resolution of GASS is $16\arcmin$, spectral resolution is $0.82$~km~s$^{-1}$, and 
sensitivity is $57$~mK. The data analyzed here are from an early release of the survey which has
not been corrected for stray radiation.
The QI and QIV regions are mirror-symmetric about the Sun--Galactic centre line, where the QI region 
spans $16.9^\circ \leq \ell \leq 35.3^\circ$ and the QIV region spans $324.7^\circ \le \ell \le 343.1^\circ$. 
Both are restricted to $|b| \lesssim 20^\circ$. 

\section{The Tangent Point Sample}

Tangent points within the inner Galaxy occur where a line of sight passes closest to the Galactic 
centre. This is where the maximum permitted velocity occurs assuming pure Galactic rotation, and this
velocity is the 
terminal velocity, $V_{\mathrm{t}}$. Random motions may push a cloud's $V_{\mathrm{LSR}}$ beyond 
$V_{\mathrm{t}}$, and the difference between the cloud's $V_{\mathrm{LSR}}$ and $V_{\mathrm{t}}$ is
defined as the deviation velocity, $V_{\mathrm{dev}}\equiv V_{\mathrm{LSR}}-V_{\mathrm{t}}$.
We define the tangent point sample of clouds as those with $V_{\mathrm{dev}} \gtrsim 0$~km~s$^{-1}$ in 
QI and $V_{\mathrm{dev}} \lesssim 0$~km~s$^{-1}$ in QIV, which results in a sample of clouds whose distances
and hence physical properties can be determined. We define $V_{\mathrm{t}}$ based on 
observational determinations by McClure-Griffiths \& Dickey (in preparation), \citet{1985Clemens},
\citet{2007McClure-Griffiths}, and \citet{2006Luna}.

\section{Properties of Individual Disk-Halo Clouds}
We detect 255 disk-halo HI clouds at tangent points within QI, but only 81 within QIV. A 
summary of the properties of the clouds within both regions is presented in 
Table~\ref{tab:p2comptable}, where $T_{\mathrm{pk}}$ is peak brightness temperature, 
$\Delta v$ is FWHM of the velocity profile, $N_{\mathrm{HI}}$ is HI column density, 
 $r$ is radius, $M_{\mathrm{HI}}$ is HI mass, 
and $|z|$ is vertical distance 
from the midplane. While there are more than three times as many clouds detected in 
QI than in QIV, the properties of both samples are quite similar, 
suggesting that the clouds in both quadrants belong to the same population of clouds.

\begin{table}[!ht]
  \caption{Median Properties of Tangent Point Disk-Halo Clouds \label{tab:p2comptable}}
  \smallskip
  \begin{center}
    {\small
      \begin{tabular}{lcccccc}
        \tableline 
        \tableline
        \noalign{\smallskip} 
        & $T_{\mathrm{pk}}$ & $\Delta v$  & $N_{\mathrm{HI}}$ &  $r$ & $M_{\mathrm{HI}}$ & $|z|$\\ 
        & [K] & [km~s$^{-1}$] & [cm$^{-2}$] & [pc] & [$M_{\odot}$] & [pc]\\ 
        \noalign{\smallskip} 
        \tableline 
        \noalign{\smallskip} 
        QI & $0.5$ & $10.6$ & $1.0\times 10^{19}$ & $28$ & $700$ & $660$\\ 
        QIV & $0.5$ & $10.6$ & $1.0\times 10^{19}$ & $32$ & $630$ & $560$\\
        \noalign{\smallskip} 
        \tableline 
      \end{tabular} 
    } 
  \end{center} 
\end{table} 

\section{Properties of the Disk-Halo Cloud Population}

\subsection{Cloud--Cloud Velocity Dispersion}

\begin{figure}
  \plotone{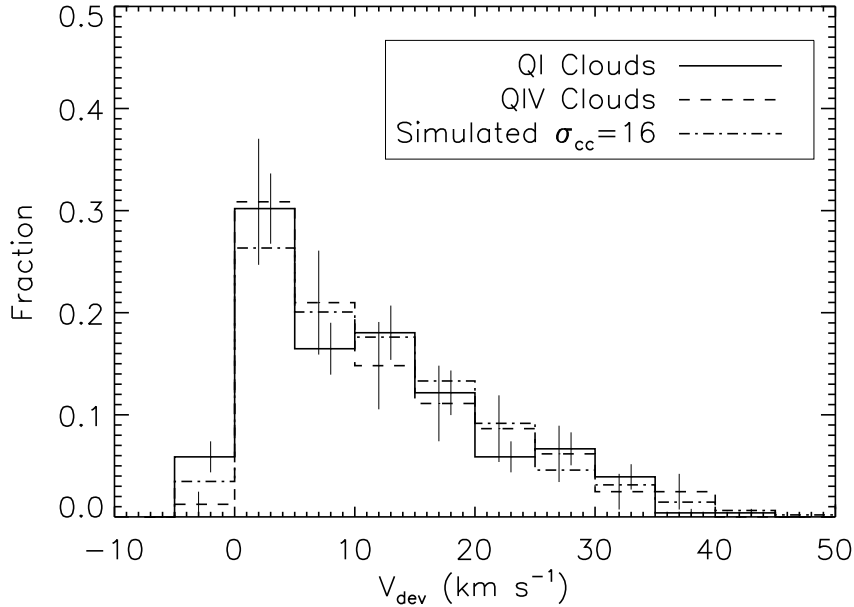}
  \caption{Distribution of observed $V_{\mathrm{dev}}$ for QI clouds (solid
            line), observed $-V_{\mathrm{dev}}$ for QIV clouds (dashed line), and
            simulated $V_{\mathrm{dev}}$ for a population of clouds with $\sigma_{cc}=16$~km~s$^{-1}$
            (dash-dotted line). 
            The QI and QIV distributions are consistent with the $\sigma_{cc}=16$~km~s$^{-1}$
            distribution.}
  \label{fig:GASS_Q14_vdev}
\end{figure}

Random motions, characterized by a cloud--cloud velocity dispersion ($\sigma_{cc}$),
can increase the $V_{\mathrm{LSR}}$ of a cloud that is located near a tangent point to
values beyond $V_{\mathrm{t}}$. We can determine the magnitude of these random motions based on the
deviation velocity distribution of the tangent point cloud samples. The observed $V_{\mathrm{dev}}$
distributions are presented in Figure \ref{fig:GASS_Q14_vdev}. The steep decline towards larger $V_{\mathrm{dev}}$
shows that the clouds' motions are governed by Galactic rotation. 
The distributions have a K-S probability of $76\%$ of being drawn from the same population, 
suggesting the QI and QIV clouds have identical $\sigma_{cc}$. Both
populations have distributions consistent with those of simulated populations that have a random 
velocity component derived from a Gaussian with dispersion $\sigma_{cc}=16$~km~s$^{-1}$. 

\subsection{Vertical Distribution}

\begin{figure}
  \plotone{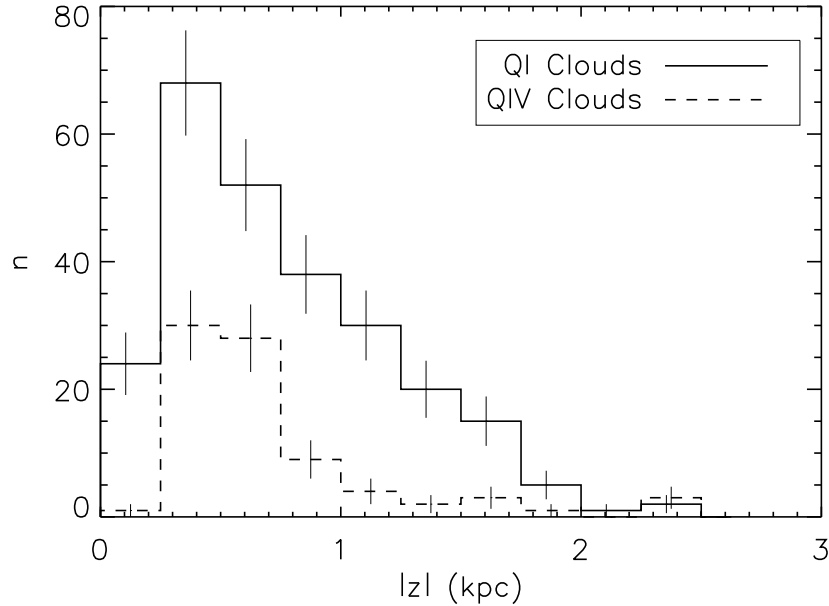}
  \caption{Distribution of observed $|z|$ for the QI (solid line) and QIV (dashed line) clouds. There are 
    significantly more clouds at most $|z|$ in QI than QIV and the scale height is twice as large.}
  \label{fig:zhistcompobs}
\end{figure}

More clouds are detected at most heights within QI than QIV, as can be seen from their
observed vertical distributions (Figure~\ref{fig:zhistcompobs}).
The decrease in the number of clouds at low $|z|$ is due to confusion, as it is increasingly 
difficult to detect clouds at lower $|z|$ and lower $|V_{\mathrm{dev}}|$. The
vertical distribution of clouds in QI is best represented by an exponential with $h=800$~pc, while in 
QIV the scale height is only 
$h=400$~pc. These values are not in agreement, and as $\sigma_{cc}$ is similar in both quadrants, 
the scale height of the clouds is not linked to the cloud--cloud velocity dispersion;  even if 
$\sigma_{cc}=\sigma_{z}$, this would only propel clouds to $h<100$~pc within the mass model of
 \citet{2007Kalberla}.

\subsection{Longitude Distribution}

\begin{figure}
  \plotone{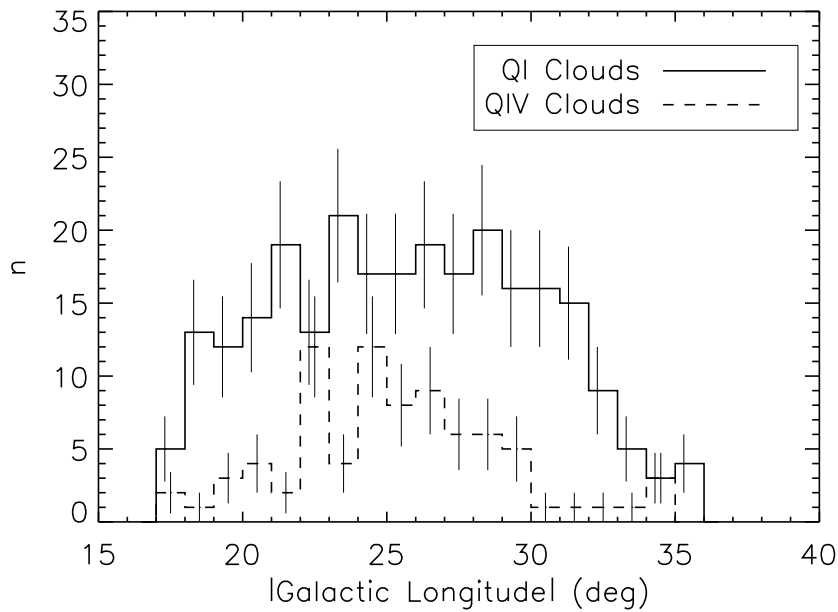}
  \caption{Observed longitude distributions of the QI (solid line) and
    QIV (dashed line) regions. More clouds are seen in QI at all $\ell$ than in QIV. 
    The distribution is more uniform in QI while it is peaked around $\ell\sim25^\circ$ in QIV.}
          \label{fig:radsurfdens}
\end{figure}

The observed longitude distributions of clouds in the QI and QIV regions are shown in 
Figure~\ref{fig:radsurfdens}. 
Not only are there more clouds at all $|\ell|$ in the QI region, but they are more uniformly 
distributed compared to the peaked QIV distribution at $\ell\sim 25^\circ$. The radial 
distribution is directly related to the longitude distribution; $R\equiv R_0\sin\ell$ at the tangent 
point, where $R_0$ is the radius of the
solar circle. The radial distribution of the QI sample is also more uniform compared to QIV, where instead
the distribution peaks and declines rapidly at $R>4.2$~kpc.

\section{Implications for the Disk-Halo Clouds}

\begin{figure}
  \plotone{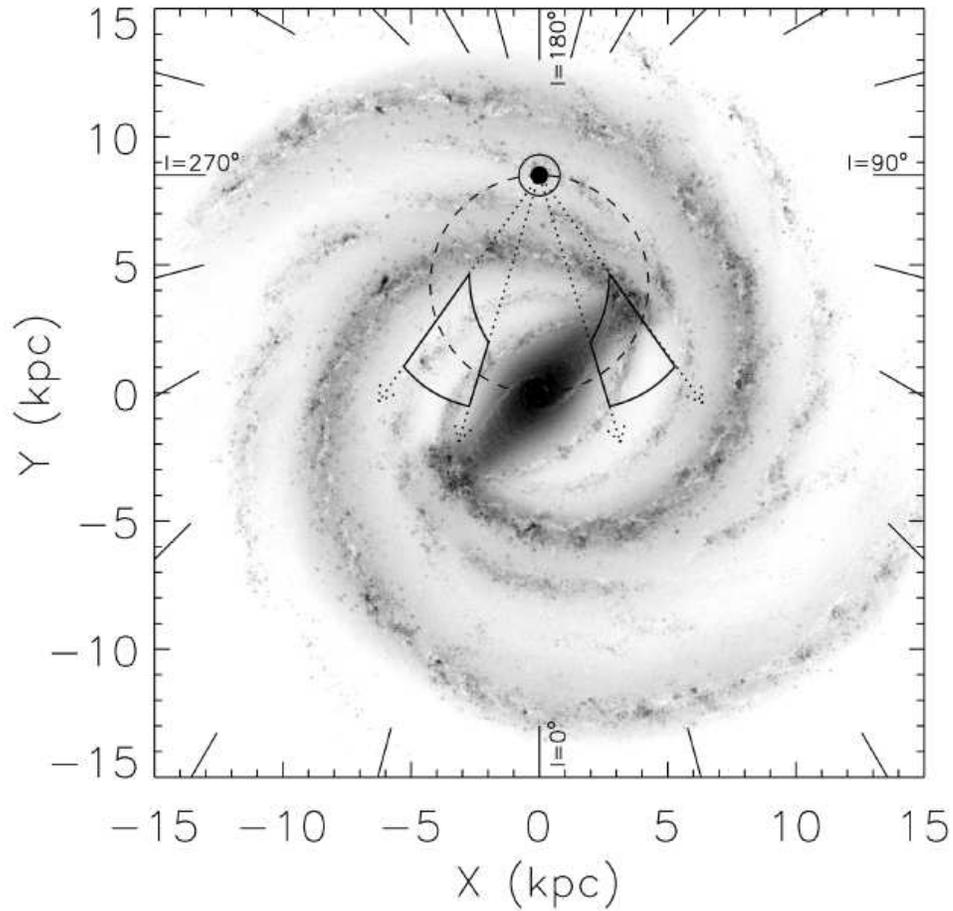}
  \caption{Figure from FLMG showing an artist's conception of the Milky Way.
    The solid lines enclose the QI and QIV regions, where QI covers the region where the near-end of the Galactic bar
    merges with a major spiral arm and QIV covers only a minor arm. The artist's conception
    image is from NASA/JPL-Caltech/R. Hurt (SSC-Caltech).}
  \label{fig:GLIMPSE_image}
\end{figure}

While the cloud--cloud velocity dispersions in both quadrants are consistent, the number of clouds 
detected, along with the vertical and radial distributions, differ dramatically. 
We have searched for possible systematic effects that might account for the asymmetry but have found none 
of any significance. These asymmetries imply that the clouds are linked to Galactic 
structure and events occurring in the disk. We have overlaid the QI and QIV regions on what we believe to be
the most appropriate representation of the large-scale Galactic structure to date, which includes recent 
results from the 
Galactic Legacy Infrared Mid-Plane Survey Extraordinaire (GLIMPSE; \citealt{2003Benjamin}) on the 
location of the bar and spiral arms (\citealt{2005Benjamin}; see Figure~\ref{fig:GLIMPSE_image}).
The solid lines enclose the QI and QIV regions, with a line of sight extent equivalent to $\sim1\sigma_{cc}$ 
volume around the tangent point. A striking asymmetry within the Galaxy that is immediately apparent is that
the QIV region contains only a minor arm while the QI region contains the merging of the near-end of the bar 
and a major spiral arm. 

The distribution of disk-halo clouds mirrors the spiral structure of the Milky Way, suggesting 
a correlation with star formation. 
The spatial relation between extraplanar gas to star formation activity in some external galaxies
further supports this hypothesis \citep{2005Barbieri}. 
More clouds occur in a region of the Galaxy including the near-end of 
the bar and a major spiral arm, while there are fewer clouds in a region including only a minor arm. The number of
clouds likely correlates with the level of star formation activity, and the different scale heights may 
represent different evolutionary stages or varying levels of star formation.
The clouds are likely formed by gas that is lifted by superbubbles and stellar feedback, and some may 
result from fragmenting supershells. Supershells have much larger timescales than star-forming regions 
($\sim 20$--$30$~Myr versus $\sim0.1$~Myr; \citealt{2004deAvillez, 2006McClure-Griffiths, 2007Prescott}), 
and so clouds may not be near regions of current star formation.
If clouds are lifted by events within the disk, large
cloud--cloud velocity dispersions are not required to produce the derived scale heights of the disk-halo
clouds. 

Disk-halo clouds are abundant within the Milky Way, and 
it is likely that many clouds would be detected at other longitudes corresponding with spiral features.

\acknowledgements We thank B.~Saxton for his help creating Figure \ref{fig:GLIMPSE_image}. 
HAF thanks 
the National Radio Astronomy Observatory for support under its Graduate Student Internship Program. 
The National Radio Astronomy Observatory is operated by Associated Universities, Inc., under a cooperative agreement 
with the National Science Foundation. The Parkes Radio Telescope is part of the Australia Telescope 
which is funded by the Commonwealth of Australia for operation as a National Facility managed by CSIRO.

\bibliography{Ford_Alyson}

\begin{thebibliography}{}
\expandafter\ifx\csname natexlab\endcsname\relax\def\natexlab#1{#1}\fi
\expandafter\ifx\csname url\endcsname\relax
  \def\url#1{\texttt{#1}}\fi
\expandafter\ifx\csname urlprefix\endcsname\relax\def\urlprefix{URL }\fi
\providecommand{\eprint}[2][]{\url{#2}}

\bibitem[{{Audit} \& {Hennebelle}(2005)}]{2005Audit}
{Audit}, E., \& {Hennebelle}, P. 2005, \aap, 433, 1

\bibitem[{{Barbieri} et~al.(2005){Barbieri}, {Fraternali}, {Oosterloo},
  {Bertin}, {Boomsma}, \& {Sancisi}}]{2005Barbieri}
{Barbieri}, C.~V., {Fraternali}, F., {Oosterloo}, T., {Bertin}, G., {Boomsma},
  R., \& {Sancisi}, R. 2005, \aap, 439, 947

\bibitem[{{Benjamin} et~al.(2003){Benjamin}, {Churchwell}, {Babler}, {Bania},
  {Clemens}, {Cohen}, {Dickey}, {Indebetouw}, {Jackson}, {Kobulnicky},
  {Lazarian}, {Marston}, {Mathis}, {Meade}, {Seager}, {Stolovy}, {Watson},
  {Whitney}, {Wolff}, \& {Wolfire}}]{2003Benjamin}
{Benjamin}, R.~A., {Churchwell}, E., {Babler}, B.~L., {Bania}, T.~M.,
  {Clemens}, D.~P., {Cohen}, M., {Dickey}, J.~M., {Indebetouw}, R., {Jackson},
  J.~M., {Kobulnicky}, H.~A., {Lazarian}, A., {Marston}, A.~P., {Mathis},
  J.~S., {Meade}, M.~R., {Seager}, S., {Stolovy}, S.~R., {Watson}, C.,
  {Whitney}, B.~A., {Wolff}, M.~J., \& {Wolfire}, M.~G. 2003, \pasp, 115, 953

\bibitem[{{Benjamin} et~al.(2005){Benjamin}, {Churchwell}, {Babler},
  {Indebetouw}, {Meade}, {Whitney}, {Watson}, {Wolfire}, {Wolff}, {Ignace},
  {Bania}, {Bracker}, {Clemens}, {Chomiuk}, {Cohen}, {Dickey}, {Jackson},
  {Kobulnicky}, {Mercer}, {Mathis}, {Stolovy}, \& {Uzpen}}]{2005Benjamin}
{Benjamin}, R.~A., {Churchwell}, E., {Babler}, B.~L., {Indebetouw}, R.,
  {Meade}, M.~R., {Whitney}, B.~A., {Watson}, C., {Wolfire}, M.~G., {Wolff},
  M.~J., {Ignace}, R., {Bania}, T.~M., {Bracker}, S., {Clemens}, D.~P.,
  {Chomiuk}, L., {Cohen}, M., {Dickey}, J.~M., {Jackson}, J.~M., {Kobulnicky},
  H.~A., {Mercer}, E.~P., {Mathis}, J.~S., {Stolovy}, S.~R., \& {Uzpen}, B.
  2005, \apjl, 630, L149

\bibitem[{{Bregman}(1980)}]{1980Bregman}
{Bregman}, J.~N. 1980, \apj, 236, 577

\bibitem[{{Clemens}(1985)}]{1985Clemens}
{Clemens}, D.~P. 1985, \apj, 295, 422

\bibitem[{{de Avillez} \& {Breitschwerdt}(2004)}]{2004deAvillez}
{de Avillez}, M.~A., \& {Breitschwerdt}, D. 2004, \aap, 425, 899

\bibitem[{{Dedes} \& {Kalberla}(2010)}]{2010Dedes}
{Dedes}, L., \& {Kalberla}, P.~W.~M. 2010, \aap, 509, 60

\bibitem[{{Ford} et~al.(2010){Ford}, {Lockman}, \&
  {McClure-Griffiths}}]{2010Ford}
{Ford}, H.~A., {Lockman}, F.~J., \& {McClure-Griffiths}, N.~M. 2010, \apj, 722,
  367

\bibitem[{{Ford} et~al.(2008){Ford}, {McClure-Griffiths}, {Lockman}, {Bailin},
  {Calabretta}, {Kalberla}, {Murphy}, \& {Pisano}}]{2008Ford}
{Ford}, H.~A., {McClure-Griffiths}, N.~M., {Lockman}, F.~J., {Bailin}, J.,
  {Calabretta}, M.~R., {Kalberla}, P.~M.~W., {Murphy}, T., \& {Pisano}, D.~J.
  2008, \apj, 688, 290

\bibitem[{{Houck} \& {Bregman}(1990)}]{1990Houck}
{Houck}, J.~C., \& {Bregman}, J.~N. 1990, \apj, 352, 506

\bibitem[{{Kalberla} et~al.(2007){Kalberla}, {Dedes}, {Kerp}, \&
  {Haud}}]{2007Kalberla}
{Kalberla}, P.~M.~W., {Dedes}, L., {Kerp}, J., \& {Haud}, U. 2007, \aap, 469,
  511

\bibitem[{{Lockman}(1984)}]{1984Lockman}
{Lockman}, F.~J. 1984, \apj, 283, 90

\bibitem[{{Lockman}(2002)}]{2002Lockman}
--- 2002, \apj, 580, L47

\bibitem[{{Luna} et~al.(2006){Luna}, {Bronfman}, {Carrasco}, \&
  {May}}]{2006Luna}
{Luna}, A., {Bronfman}, L., {Carrasco}, L., \& {May}, J. 2006, \apj, 641, 938

\bibitem[{{McClure-Griffiths} \& {Dickey}(2007)}]{2007McClure-Griffiths}
{McClure-Griffiths}, N.~M., \& {Dickey}, J.~M. 2007, \apj, 671, 427

\bibitem[{{McClure-Griffiths} et~al.(2006){McClure-Griffiths}, {Ford},
  {Pisano}, {Gibson}, {Staveley-Smith}, {Calabretta}, {Dedes}, \&
  {Kalberla}}]{2006McClure-Griffiths}
{McClure-Griffiths}, N.~M., {Ford}, A., {Pisano}, D.~J., {Gibson}, B.~K.,
  {Staveley-Smith}, L., {Calabretta}, M.~R., {Dedes}, L., \& {Kalberla},
  P.~M.~W. 2006, \apj, 638, 196

\bibitem[{{McClure-Griffiths} et~al.(2009){McClure-Griffiths}, {Pisano},
  {Calabretta}, {Ford}, {Lockman}, {Staveley-Smith}, {Kalberla}, {Bailin},
  {Dedes}, {Janowiecki}, {Gibson}, {Murphy}, {Nakanishi}, \&
  {Newton-McGee}}]{2009McClure-Griffiths}
{McClure-Griffiths}, N.~M., {Pisano}, D.~J., {Calabretta}, M.~R., {Ford},
  H.~A., {Lockman}, F.~J., {Staveley-Smith}, L., {Kalberla}, P.~M.~W.,
  {Bailin}, J., {Dedes}, L., {Janowiecki}, S., {Gibson}, B.~K., {Murphy}, T.,
  {Nakanishi}, H., \& {Newton-McGee}, K. 2009, \apjs, 181, 398

\bibitem[{{Prata}(1964)}]{1964Prata}
{Prata}, S.~W. 1964, Bull. Astron. Inst. Netherlands, 17, 511

\bibitem[{{Prescott} et~al.(2007){Prescott}, {Kennicutt}, {Bendo}, {Buckalew},
  {Calzetti}, {Engelbracht}, {Gordon}, {Hollenbach}, {Lee}, {Moustakas},
  {Dale}, {Helou}, {Jarrett}, {Murphy}, {Smith}, {Akiyama}, \&
  {Sosey}}]{2007Prescott}
{Prescott}, M.~K.~M., {Kennicutt}, R.~C., Jr., {Bendo}, G.~J., {Buckalew},
  B.~A., {Calzetti}, D., {Engelbracht}, C.~W., {Gordon}, K.~D., {Hollenbach},
  D.~J., {Lee}, J.~C., {Moustakas}, J., {Dale}, D.~A., {Helou}, G., {Jarrett},
  T.~H., {Murphy}, E.~J., {Smith}, J., {Akiyama}, S., \& {Sosey}, M.~L. 2007,
  \apj, 668, 182

\bibitem[{{Shapiro} \& {Field}(1976)}]{1976Shapiro}
{Shapiro}, P.~R., \& {Field}, G.~B. 1976, \apj, 205, 762

\bibitem[{{Simonson}(1971)}]{1971Simonson}
{Simonson}, S.~C., III 1971, \aap, 12, 136

\bibitem[{{Spitoni} et~al.(2008){Spitoni}, {Recchi}, \&
  {Matteucci}}]{2008Spitoni}
{Spitoni}, E., {Recchi}, S., \& {Matteucci}, F. 2008, \aap, 484, 743

\bibitem[{{Stanimirovi{\'c}} et~al.(2006){Stanimirovi{\'c}}, {Putman},
  {Heiles}, {Peek}, {Goldsmith}, {Koo}, {Kr{\v c}o}, {Lee}, {Mock}, {Muller},
  {Pandian}, {Parsons}, {Tang}, \& {Werthimer}}]{2006Stanimirovic}
{Stanimirovi{\'c}}, S., {Putman}, M., {Heiles}, C., {Peek}, J.~E.~G.,
  {Goldsmith}, P.~F., {Koo}, B.-C., {Kr{\v c}o}, M., {Lee}, J.-J., {Mock}, J.,
  {Muller}, E., {Pandian}, J.~D., {Parsons}, A., {Tang}, Y., \& {Werthimer}, D.
  2006, \apj, 653, 1210

\bibitem[{{Stil} et~al.(2006){Stil}, {Lockman}, {Taylor}, {Dickey}, {Kavars},
  {Martin}, {Rothwell}, {Boothroyd}, \& {McClure-Griffiths}}]{2006Stil}
{Stil}, J.~M., {Lockman}, F.~J., {Taylor}, A.~R., {Dickey}, J.~M., {Kavars},
  D.~W., {Martin}, P.~G., {Rothwell}, T.~A., {Boothroyd}, A.~I., \&
  {McClure-Griffiths}, N.~M. 2006, \apj, 637, 366

\end{thebibliography}

\end{document}